\documentclass[prd,twocolumn,a4paper,superscriptaddress]{revtex4}
\usepackage{graphicx,amssymb}
\begin{document}
%My commands
\newcommand{\be}{\begin{equation}}
\newcommand{\ee}{\end{equation}}
\newcommand{\bq}{\begin{eqnarray}}
\newcommand{\eq}{\end{eqnarray}}
\newcommand{\bsq}{\begin{subequations}}
\newcommand{\esq}{\end{subequations}}
\newcommand{\bc}{\begin{center}}
\newcommand{\ec}{\end{center}}
\newcommand {\R}{{\mathcal R}}
\newcommand{\al}{\alpha}
\newcommand\lsim{\mathrel{\rlap{\lower4pt\hbox{\hskip1pt$\sim$}}
    \raise1pt\hbox{$<$}}}
\newcommand\gsim{\mathrel{\rlap{\lower4pt\hbox{\hskip1pt$\sim$}}
    \raise1pt\hbox{$>$}}}

\title{Scaling Properties of Domain Wall Networks}
\author{A.M.M. Leite}
\email{up080322016@alunos.fc.up.pt}
\affiliation{Centro de Astrof\'{\i}sica, Universidade do Porto, Rua das Estrelas, 4150-762 Porto, Portugal}
\affiliation{Faculdade de Ci\^encias, Universidade do Porto, Rua do Campo Alegre 687, 4169-007 Porto, Portugal}
\author{C.J.A.P. Martins}
\email[Electronic address: ]{Carlos.Martins@astro.up.pt}
\affiliation{Centro de Astrof\'{\i}sica, Universidade do Porto, Rua das Estrelas, 4150-762 Porto, Portugal}
\affiliation{Department of Applied Mathematics and Theoretical Physics, Centre for Mathematical Sciences,\\ University of Cambridge, Wilberforce Road, Cambridge CB3 0WA, United Kingdom}
\date{14 October 2011}

\begin{abstract}
We revisit the cosmological evolution of domain wall networks, taking advantage of recent improvements in computing power. We carry out high-resolution field theory simulations in two, three and four spatial dimensions to study the effects of dimensionality and damping on the evolution of the network. Our results are consistent with the expected scale-invariant evolution of the network, which suggests that previous hints of deviations from this behavior may have been due to the limited dynamical range of those simulations. We also use the results of very large ($1024^3$) simulations in three cosmological epochs to provide a calibration for the velocity-dependent one-scale model for domain walls: we numerically determine the two free model parameters to have the values $c_w=0.5\pm0.2$ and $k_w=1.1\pm0.3$.
\end{abstract}
\maketitle

\section{\label{intr}Introduction}

Phase transitions that are thought to have happened in the early universe have a number of inevitable consequences, the most interesting of which is the formation of topological defects \cite{Kibble,Book}. The literature on the subject has (for good reasons) focused on cosmic strings, but other defects can be of interest too. Domain walls, being the simplest defect (they can be described by a single scalar field) provide a simple testbed where one can study how several physical mechanisms influence defect evolution, and this knowledge can then be applied to other defects. This is despite the fact that the observational roles of domain walls are very tightly constrained: the Zel'dovich bound \cite{Zeldovich} rules them out if their symmetry breaking scale is $\eta\ge 1MeV$, and the bound is even tighter for wall networks with junctions \cite{Junctions}. 

Here we take advantage of the continuous improvements in computing power to carry out a large set of high-resolution simulations of domain walls in two, three and four spatial dimensions, using the standard Press-Ryden-Spergel (PRS) algorithm \cite{Press}. While the 3D simulations are of obvious cosmological relevance, the 2D ones have the advantage of allowing for a much larger dynamical range, and the 4D ones may be relevant to some brane world scenarios \cite{tye0,Brax,Matsuda,Barnaby}. We'll only consider the simplest domain wall model, thus neglecting scenarios where domain walls have junctions.

One can also describe the broad macroscopic properties of these networks by an analytic model, in the same spirit of the model of Martins and Shellard for cosmic strings \cite{ms1a,ms1b,extend}. The large-scale features of the network are described by a lengthscale (or correlation length) $L$ and a microscopically averaged (root-mean-squared) velocity $v$. Although this has the advantages of tractability and conceptual simplicity, these come with a price: in going from the microphysics to the averaged evolution equations one is forced to introduce phenomenological parameters which parametrize our ignorance about certain dynamical mechanisms, and the only way to determine the correct values of these parameters is by referring to numerical simulations to calibrate them.

The evolution of domain walls networks has been previously studied, numerically, by a number of different authors \cite{Press,Coulson,Larsson,Fossils,Garagounis,MY1,MY2,Lalak}, who usually found some hints for deviations to the scale-invariant evolution. Since there are good reasons to expect this scale-invariant attractor \cite{Hindmarsh,MY2}, one may ask whether these deviations point to the presence of physical mechanisms not accounted for in analytic descriptions or if they are simply a consequence of the limited dynamical range of numerical simulations. While the present work cannot completely resolve this issue, we believe that it provides support for the second alternative.

In what follows we briefly describe the PRS algorithm used in numerical simulations, as well as the analytic model we will use to describe the domain wall networks. We then proceed to present the main results of our simulations, and finally bring the two approaches together by comparing the simulation results to the analytic model predictions, thereby providing a calibration of the free parameters of the latter. We'll conclude with brief comments on the cosmological implications of our results and on possible follow-up work. Throughout the paper we shall use fundamental units, in which $c=\hbar=1$.

\section{\label{prs}Domain Walls and the PRS Algorithm}

We'll be interested in domain wall networks in flat homogeneous and isotropic Friedmann-Robertson-Walker (FRW) universes. A scalar field $\phi$ with Lagrangian density
\begin{equation}
\mathcal{L}={\frac{1}{2}}\phi_{,\alpha}\phi^{,\alpha}-V(\phi)\,,
\label{action1}
\end{equation}
where we will take $V(\phi)$ to be a $\phi^{4}$ potential with two degenerate minima, such as
\begin{equation}
V(\phi)=V_{0}\left({\frac{\phi^{2}}{\phi_{0}^{2}}}-1\right)^{2}\,,
\label{potential}
\end{equation}
will have domain wall solutions. By the standard variational methods we obtain the field equation of motion (written in terms of physical time $t$) 
\begin{equation}
{\frac{{\partial^{2}\phi}}{\partial 
t^{2}}}+3H{\frac{{\partial\phi}}{\partial 
t}}-\nabla^{2}\phi=-{\frac{{\partial 
V}}{\partial\phi}}\,.\label{dynamics}
\end{equation}
where $\nabla$ is the Laplacian in physical coordinates, $H=a^{-1}(da/dt)$ is the Hubble parameter and $a$ is the scale factor, which we assume to vary as $a\propto t^\lambda$. In what follows we will be interested in comparing the network evolution  in several different cosmological epochs, in particular the radiation era ($\lambda=1/2$), the matter era ($\lambda=2/3$) and a fast expansion era ($\lambda=4/5$).

We then apply the procedure of Press, Ryden and Spergel \cite{Press}, modifying the equations of motion in such a way that the thickness of the domain walls is fixed in co-moving coordinates. One expects that this will have a small impact on the large scale dynamics of the domain walls, since a wall's integrated surface density (and surface tension) are independent of its thickness. In particular, this assumption should not affect the presence or absence of a scaling solution \cite{Press}, provided one uses a minimum thickness \cite{MY1}---we will briefly revisit this issue below. (For a detailed discussion of analogous issues in the context of cosmic strings see \cite{Moore}.)

In the PRS method, equation (\ref{dynamics}) becomes: 
\begin{equation}
{\frac{{\partial^{2}\phi}}{\partial\eta^{2}}}+\alpha\left(\frac{d\ln 
a}{d\ln\eta}\right){\frac{{\partial\phi}}{\partial\eta}}-{\nabla}^{2}\phi=
-a^{\beta}{\frac{{\partial 
V}}{\partial\phi}}\,.\label{dynamics2}
\end{equation}
where $\eta$ is the conformal time and $\alpha$ and $\beta$ are constants: $\beta=0$ is used in order to have constant co-moving thickness and $\alpha=3$ is chosen (in 3D, see \cite{Sousa} for an argument in other dimensions) to require that the momentum conservation law of the wall evolution in an expanding universe is maintained \cite{Press}. In fact we have simulated networks with various values of the damping coefficient $\alpha$.

Equation (\ref{dynamics2}) is then integrated using a standard finite-difference scheme. We assume the initial value of $\phi$ to be a random variable between $-\phi_{0}$ and $+\phi_{0}$ and the initial value of $\partial\phi/\partial\eta$ to be zero. This will lead to large energy gradients in the early timesteps of the simulation, and therefore the network will need some time (which is proportional to the wall thickness) to wash away these initial conditions. The conformal time evolution of the co-moving correlation length of the network $\xi_c$ (specifically $A/V\propto \xi_{c}^{-1}$, $A$ being the comoving area of the walls) and the wall velocities (specifically $\gamma v$, where $\gamma$ is the Lorentz factor) are directly measured from the simulations, using techniques previously described in \cite{MY2}. However we now used a newly parallelized version of the code, optimized for the Altix UV1000 architecture of the COSMOS Consortium's \textit{Universe} supercomputer.

\section{\label{model}The one-scale model}

The way to analytically model defect networks is to start from their microscopic equations of motion (the Nambu-Goto equations, in the case of strings) and carry out a statistical average, under the assumption that the defects are randomly distributed at large enough scales. This leads to a macroscopic energy evolution equation (which one can equivalently write as an equation for the network's correlation length or for another suitable characteristic lengthscale) and an equation for the network's root-mean squared (RMS) velocity.

These evolution equations provide a 'thermodynamical' description of the network, in the same sense that the microscopic equations provide a statistical physics one. The more subtle part of this procedure is that of the addition of terms in these equations to account for defect interactions and energy losses. Such terms must be added in a phenomenological way, and for their calibration one must resort to numerical simulations.

For cosmic strings, this procedure leads to the so-called velocity-dependent one-scale (VOS) model \cite{ms1a,ms1b,extend}, which has been well-tested against simulations and is used for predicting CMB signals of string networks. One can follow an analogous procedure both for the case of monopoles \cite{Monopoles} and for domain walls. This latter case has been described in \cite{MY2}, which also provided a simple (qualitative) calibration; later in this article we will revisit this issue and provide a more quantitative one.

The evolution equation for the characteristic wall lengthscale $L$ (which is related to the wall density $\rho_w$ via $L=\sigma/\rho_w$, where $\sigma$ is the domain wall energy per unit area) and their RMS velocity $v$, are as follows
\begin{equation}
\frac{dL}{dt}=(1+3v^2)HL+c_wv\,
\label{rhoevoldw1}
\end{equation}
\begin{equation}
\frac{dv}{dt}=(1-v^2)\left(\frac{k_w}{L}-3Hv\right)\,;
\label{vevoldw1}
\end{equation}
the latter equation was first obtained, using a different approximation,  in \cite{kawano}. Here $c_w$ and $k_w$ are the free parameters, to be calibrated using simulations: the former quantifies energy losses, while the latter quantifies the (curvature-related) forces acting on the walls. At least to a first approximation, these are expected to be constant. Moreover, in the context of the VOS model the characteristic lengthscale $L$ can further be identified with the physical correlation length $\xi_{phys}$. The comoving version of this was defined in the previous section; the two are related through 
\begin{equation}
\xi_{phys}=a\xi_c\,,
\end{equation}
and we are therefore assuming that $\xi_{phys}\equiv L$. We will use the two terms interchangeably for the rest of this paper. Note that the scaling exponents of the two correlation lengths, relative to their respective times, are different: if
\begin{equation}
\xi_c\propto \eta^{1-\delta}\,,
\end{equation}
then
\begin{equation}
\xi_{phys}\propto t^{1-\delta(1-\lambda)}\,,
\end{equation}
for an expansion rate $\lambda$ defined as before.

If one neglects the effect of the energy density in the domain walls on the background (specifically, on the Friedmann equations)---which is the relevant context for our numerical simulations---it is easy to see that, just as for cosmic string networks, the attractor solution to the evolution equations (\ref{rhoevoldw1},\ref{vevoldw1}) corresponds to a linear scaling solution
\begin{equation}
L=\epsilon t\,, \qquad v=const\,.
\label{defscaling}
\end{equation}
Assuming that the scale factor behaves as $a \propto t^\lambda$ the detailed form of the above linear scaling constants is
\begin{equation}
\epsilon^2=\frac{k_w(k_w+c_w)}{3 \lambda (1-\lambda)}\,
\label{scaling1}
\end{equation}
\begin{equation}
v^2=\frac{1-\lambda}{3\lambda}\frac{k_w}{k_w+c_w}\,.
\label{scaling2}
\end{equation}
As in the case of cosmic strings \cite{nonint}, an energy loss mechanism (that is, a non-zero $c_w$) may not be needed in order to have linear scaling. In the $c_w\to 0$ limit one finds that for
$\alpha>1/4$ a linear scaling solution is always possible. Therefore, a linear scaling solution for domain walls can always exist in both the matter and the radiation eras, which shows that having non-standard (that is non-intercommuting) domain walls is by no means sufficient to ensure a frustrated wall network.

In passing, we note that the cosmological linear scaling solutions for walls imply that the wall density grows relative to the background density, and will eventually become dominant (unless some mechanism like a subsequent phase transition were to make it decay and disappear). This is ultimately the reason for the Zel'dovich bound \cite{Zeldovich}. Moreover, since the wall density grows relative to the background, it must be included in the Einstein equations. In this case (further discussed in \cite{Zeldovich,MY2}) the domain wall network will become frozen in comoving coordinates with $L\propto a$ and the scale factor growing as $a\propto t^{2}$. Notice that this solution does not depend on $c_w$---domain wall interactions play no role here since the walls are effectively frozen.

\begin{table*}
\begin{tabular}{|c|c|c|c|c|c|c|}
\hline 
Box Size&
$\alpha$&
Fit range ($\eta$)&
$\mu$&
$\nu$&
$\xi_c/\eta$&
$\gamma v$\tabularnewline
\hline 
\hline $4096^2$ & $2.0$ & 41-2048 & $-0.95\pm0.04$ & $-0.00003\pm0.00002$  & $0.61$ & $0.37$ \tabularnewline
\hline $4096^2$ & $3.0$ & 23.5-2048 & $-0.97\pm0.05$ & $-0.00002\pm0.00002$ & $0.58$ & $0.34$ \tabularnewline
\hline $4096^2$ & $4.0$ & 21-2048 & $-0.96\pm0.04$ & $-0.00001\pm0.00002$ & $0.55$ & $0.33$ \tabularnewline
\hline
\hline $8192^2$ & $2.0$ & 12.25-4096 & $-0.98\pm0.03$ & $-0.000011\pm0.000008$ & $0.91$ & $0.40$ \tabularnewline
\hline $8192^2$ & $3.0$ & 21-4096 & $-0.97\pm0.05$ & $-0.000011\pm0.000006$ & $0.82$ & $0.32$ \tabularnewline
\hline $8192^2$ & $4.0$ & 21-4096 & $-0.96\pm0.04$ & $-0.000007\pm0.000005$ & $0.76$ & $0.29$ \tabularnewline
\hline
\hline $256^3$ & $2.0$ & 38.5-128 & $-0.94\pm0.09$ & $-0.0000\pm0.0005$ & $0.56$ & $0.44$ \tabularnewline
\hline $256^3$ & $3.0$ & 26-128 & $-0.92\pm0.05$ & $-0.0000\pm0.0003$ & $0.52$ & $0.37$ \tabularnewline
\hline $256^3$ & $4.0$ & 26-128 & $-0.94\pm0.04$ & $-0.0000\pm0.0002$ & $0.49$ & $0.33$ \tabularnewline
\hline
\hline $512^3$ & $2.0$ & 38.5-256 & $-0.96\pm0.05$ & $-0.0001\pm0.0001$ & $0.57$ & $0.43$ \tabularnewline
\hline $512^3$ & $3.0$ & 26-256 & $-0.97\pm0.04$ & $-0.0000\pm0.0001$ & $0.54$ & $0.37$ \tabularnewline
\hline $512^3$ & $4.0$ & 26-256 & $-0.96\pm0.03$ & $0.0000\pm0.0001$ & $0.50$ & $0.34$ \tabularnewline
\hline
\hline $64^4$ & $2.0$ & 21-32 & $-1.0\pm0.1$ & $0.001\pm0.002$ & $0.53$ & $0.47$ \tabularnewline
\hline $64^4$ & $3.0$ & 21-32 & $-1.0\pm0.1$ & $0.000\pm0.002$ & $0.49$ & $0.35$ \tabularnewline
\hline $64^4$ & $4.0$ & 21-32 & $-1.0\pm0.1$ & $0.001\pm0.001$ & $0.49$ & $0.35$ \tabularnewline
\hline
\hline $128^4$ & $2.0$ & 38.5-64 & $-1.0\pm0.2$ & $0.0000\pm0.0009$ & $0.60$ & $0.47$ \tabularnewline
\hline $128^4$ & $3.0$ & 21-64 & $-0.9\pm0.1$ & $0.0002\pm0.0005$ & $0.54$ & $0.42$ \tabularnewline
\hline $128^4$ & $4.0$ & 23.5-64 & $-0.9\pm0.1$ & $0.0003\pm0.0001$ & $0.54$ & $0.40$ \tabularnewline
\hline
\end{tabular}
\caption{\label{tab1}The measured scaling exponents $\mu$ and $\nu$ (with one-sigma statistical uncertainties) for $2D$, $3D$ and $4D$ numerical simulations of domain wall networks with different box sizes and damping terms ($\alpha$). The third column shows the range of the part of the simulation that was actually used in order to fit for the scaling exponent. The last two columns show the directly measured asymptotic values of $\xi_c/\eta$ and $\gamma v$, which can be related to the macroscopic parameters of the analytic model. All simulations were done in a matter-dominated era ($\lambda=2/3$),with $\beta=0$ and a constant wall thickness $W_0=10$. Each value of the scaling exponent is obtained by averaging over 25 simulations with different random initial conditions.}
\end{table*}

\begin{figure}
\includegraphics[width=3.5in]{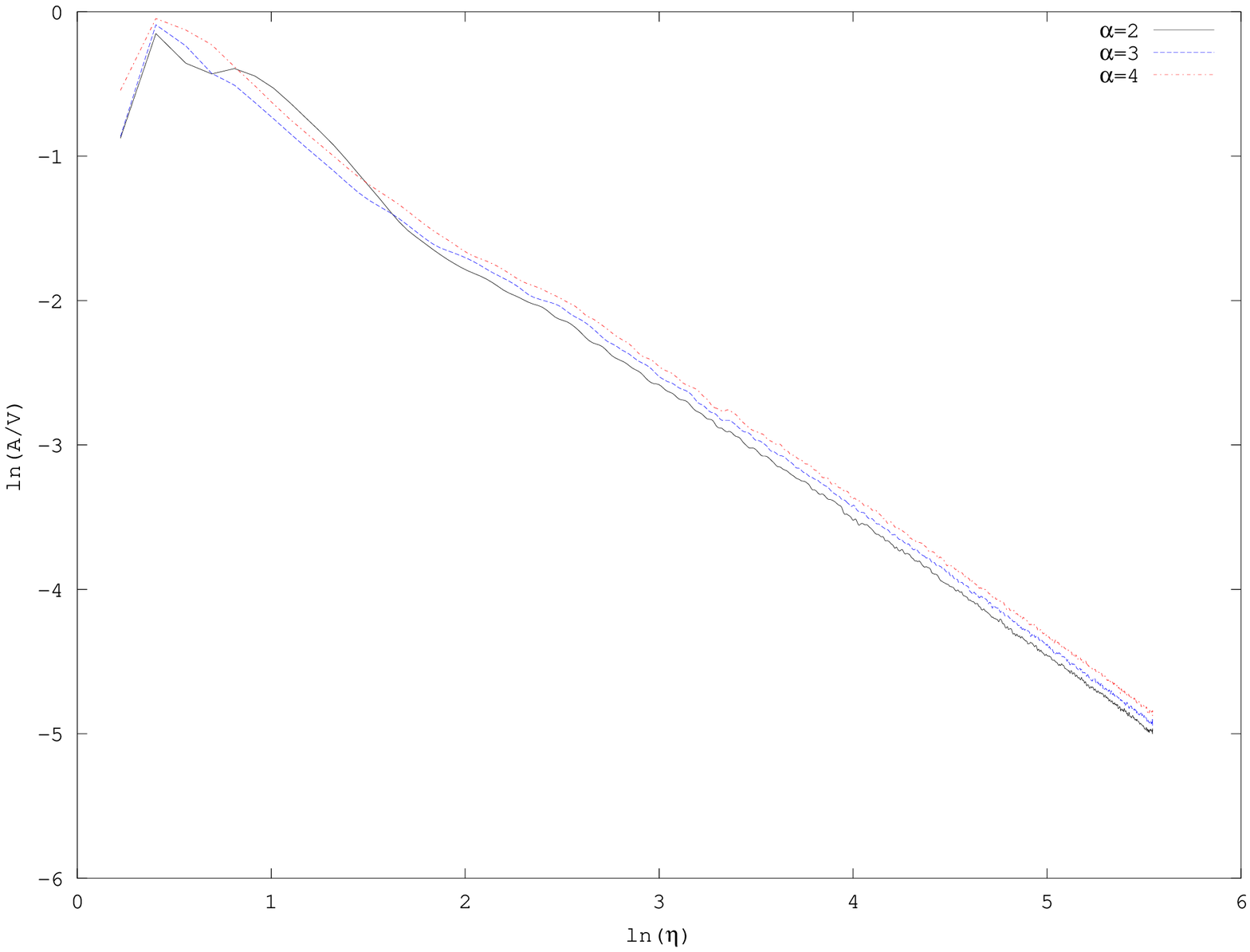}
\includegraphics[width=3.5in]{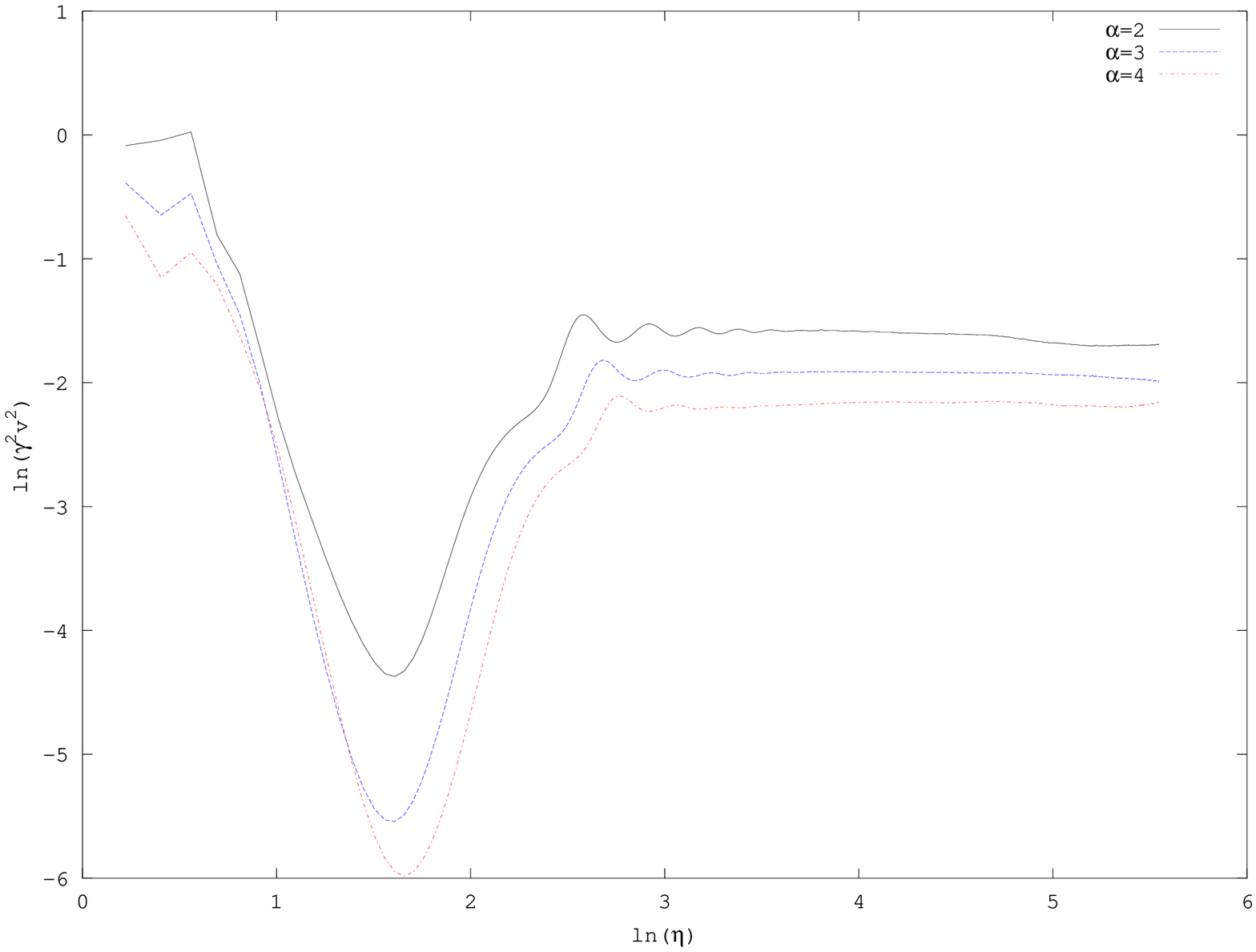}
\caption{\label{fig1}The evolution of the domain wall density ($\rho_w=A/V$) and velocity ($(\gamma v)^2$) as a function of conformal time, for the  3D $512^3$ boxes of Table \protect\ref{tab1}, corresponding to different amounts of damping, $\alpha=2,3,4$. The plotted curves are the average of 25 different simulations with different (random) initial conditions.}
\end{figure}

\begin{figure}
\includegraphics[width=3.5in]{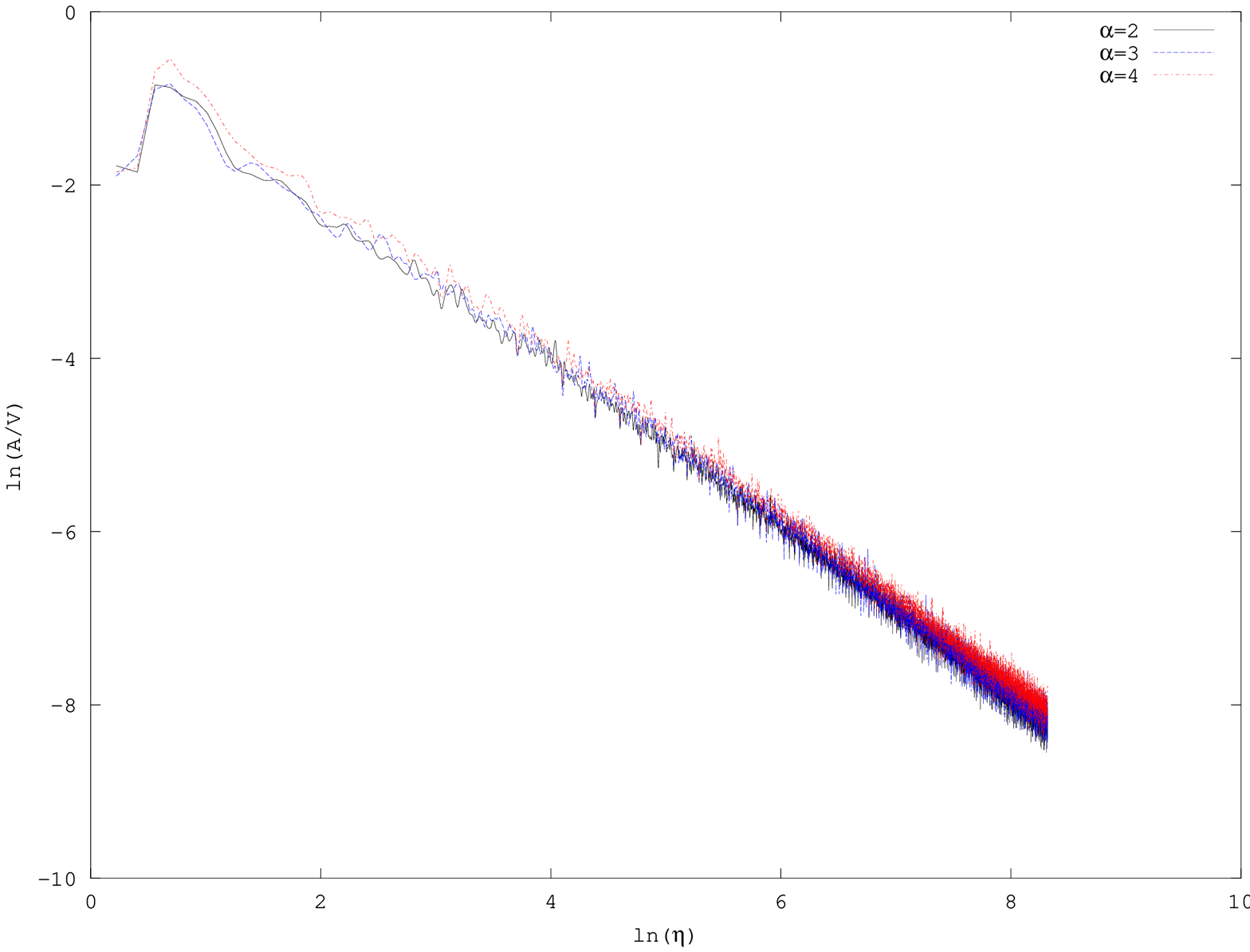}
\includegraphics[width=3.5in]{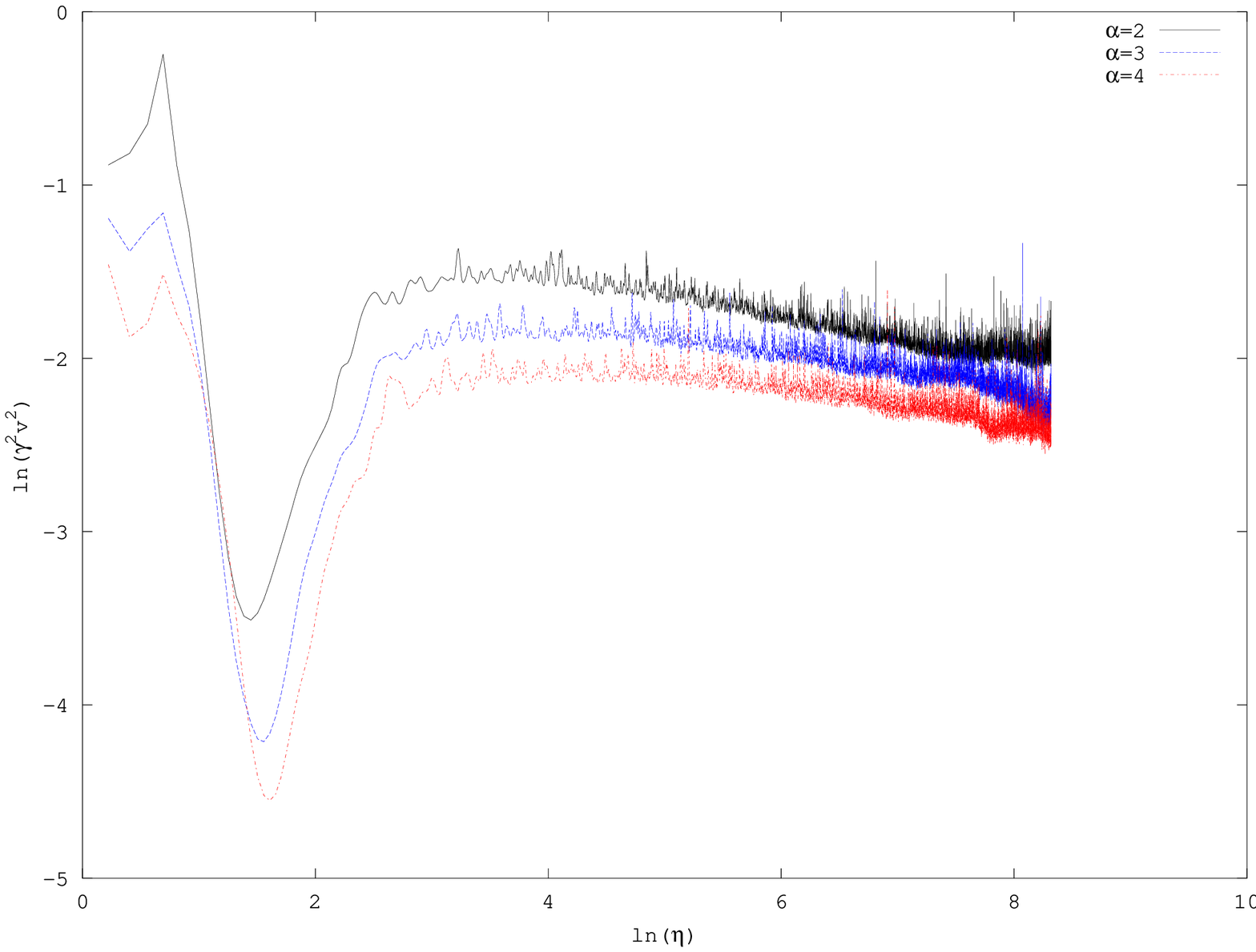}
\caption{\label{fig2}Same as Fig. \protect\ref{fig1}, for the 2D $8192^2$ simulations.}
\end{figure}

\section{\label{tests}Box size, damping and wall thickness effects}

We have started by running several sets of simulations in two, three and four spatial dimensions, varying the box size, the damping parameter $\alpha$ and the wall thickness (the typical number of points describing each wall, denoted $W_0$), as further detailed below. These---and indeed all the simulations described in this paper---were started at a conformal time $\eta_0=1$ and evolved in timesteps $\Delta\eta=0.25\eta_0$ until a conformal time equal to half the box size, and each set of simulations contains 25 runs with different (random) initial conditions. Unless otherwise stated the quoted errors are statistical errors in the ensemble of 25 runs.

Our main concern here is with a diagnostic for scaling. We looked for the best fit to the power laws
\begin{equation}
\frac{A}{V}\propto\rho_{w}\propto\frac{1}{\xi_{c}}\propto\eta^{\mu}\,,
\label{fit1}
\end{equation}
\begin{equation}
\gamma v\propto\eta^{\nu}\,;
\label{fit2}
\end{equation}
for a scale-invariant behavior, we should have $\mu=-1$ and $\nu=0$. The behavior of the scaling exponent for the network's kinetic energy (or velocity), $\nu$, has not been explored in detail in previous work. For the scaling of the correlation length, it has been suggested that this can be slightly larger than $\mu=-1$, which would imply that the network is not evolving as fast as is allowed by causality. This would presumably suggest the presence of physical mechanisms influencing the network that were not being accounted for.

One must be careful to fit only the reliable dynamical range of each simulation. As previously pointed out, the dynamics at the beginning of the simulation will be dominated by our choice of initial conditions will influence the evolution. Moreover, for the walls to be sufficiently well defined (which is certainly helpful in accurately measuring walls areas and velocities) the co-moving correlation length should be significantly larger than the wall thickness. Since we end all the simulations when the horizon becomes half the box size, the periodic boundary conditions should have no influence on our results. Our choice of the reliable period for the fit was done by inspection of each set of simulations, using these criteria \cite{Press}. 

Table \ref{tab1} shows the compared results of matter-era simulations where we varied the box size and the amount of damping (parametrized by $\alpha$) while keeping a constant wall thickness as in the PRS prescription. The key result here is that all the scaling exponents are consistent with the presence of a scale-invariant evolution of the network, corresponding to $\mu=-1$ and $\nu=0$. The constancy of the wall velocities (which had not been studied in quantitative detail by previous authors) is particularly noteworthy.

We see no hints of the deviations from this scaling behavior that have been discussed by previous authors, particularly \cite{Press} and \cite{MY1}. (On the other hand, \cite{Garagounis} finds a possible deviation in the largest of their simulations, while the smaller ones are consistent with scaling.) We therefore suggest that such hints may have been due to the limited dynamical range of simulations. Interestingly, one can observe (particularly in the 3D case) that doubling the box size tends to bring the scaling exponent $\mu$ closer to $-1$, although it has to be said that this may in part be due to the (somewhat arbitrary) choice of the dynamical range that is used for the fit. We will comment of this point again in the next section.

In addition to showing the calculated scaling exponents $\mu$ and $\nu$, the table also shows the directly measured asymptotic values of $\xi_c/\eta$ and $\gamma v$, which can be related to the macroscopic parameters of the analytic model. These are calculated from the last few timesteps of each simulation, on the assumption that by then the network has reached scaling---note that our measured scaling exponents are consistent with this assumption. Given the simplicity of this method we have chosen not to present the (statistical) error bars on these numbers, which should therefore be seen as qualitative indicators of the properties of the network. (The errors we find are typically at the ten to twenty percent level, with some dependence on the box size.) A more detailed analysis of the scaling properties will be discussed in the next section. 

These results also confirm the expectation that increasing the amount of damping leads to slower walls. One of the consequences of this is that reduces the rate of wall intercommutings and formation of closed walls. Since this is a key energy loss mechanism form the wall network, it follows that a larger damping also leads to a higher wall density, corresponding to a smaller wall separation or correlation length.

Figs. \ref{fig1} and \ref{fig2} show two examples for 2D and 3D simulations. The fact that the 2D plots show much larger fluctuations is due to the fact that the precision of our algorithm for measuring domain wall areas and velocities is somewhat dependent on dimensionality and box size. We have chosen to always use the same algorithm, described in \cite{MY2}, rather than individually optimizing it for different boxes.

\begin{table*}
\begin{tabular}{|c|c|c|c|c|c|c|}
\hline 
$\lambda$&
$W_0$&
Fit range ($\eta$)&
$\mu$&
$\nu$&
$\xi_c/\eta$&
$\gamma v$\tabularnewline
\hline 
\hline 1/2 & 10 & 38.5-256 & $-0.99\pm0.05$ & $-0.0001\pm0.0002$ & $0.60$ & $0.46$ \tabularnewline
\hline 1/2 & 50 & 188.5-256 & $-1.0\pm0.5$ & $0.0001\pm0.0008$ & $1.00$ & $0.53$ \tabularnewline
\hline 1/2 & 100 & 188.5-256 & $-1.0\pm0.4$ & $0.0003\pm0.0007$ & $0.90$ & $0.52$ \tabularnewline
\hline
\hline 4/5 & 10 & 21-256 & $-0.96\pm0.03$ & $0.00001\pm0.00005$ & $0.44$ & $0.28$ \tabularnewline
\hline 4/5 & 100 & 176-256 & $-0.9\pm0.3$ & $0.0005\pm0.0003$ & $0.67$ & $0.28$ \tabularnewline
\hline
\end{tabular}
\caption{\label{tab2}The measured scaling exponents $\mu$ and $\nu$ (with one-sigma statistical uncertainties) for $3D$ numerical simulations of domain wall networks with different expansion rates ($\lambda$) and wall thicknesses ($W_0$). The third column shows the range of the part of the simulation that was actually used in order to fit for the scaling exponent. The last two columns show the directly measured asymptotic values of $\xi_c/\eta$ and $\gamma v$, which can be related to the macroscopic parameters of the analytic model. All simulations were done in $512^3$ boxes with $\alpha=3$ and $\beta=0$. Each value of the scaling exponent is obtained by averaging over 25 simulations with different random initial conditions.}
\end{table*}

\begin{figure}
\includegraphics[width=3.5in]{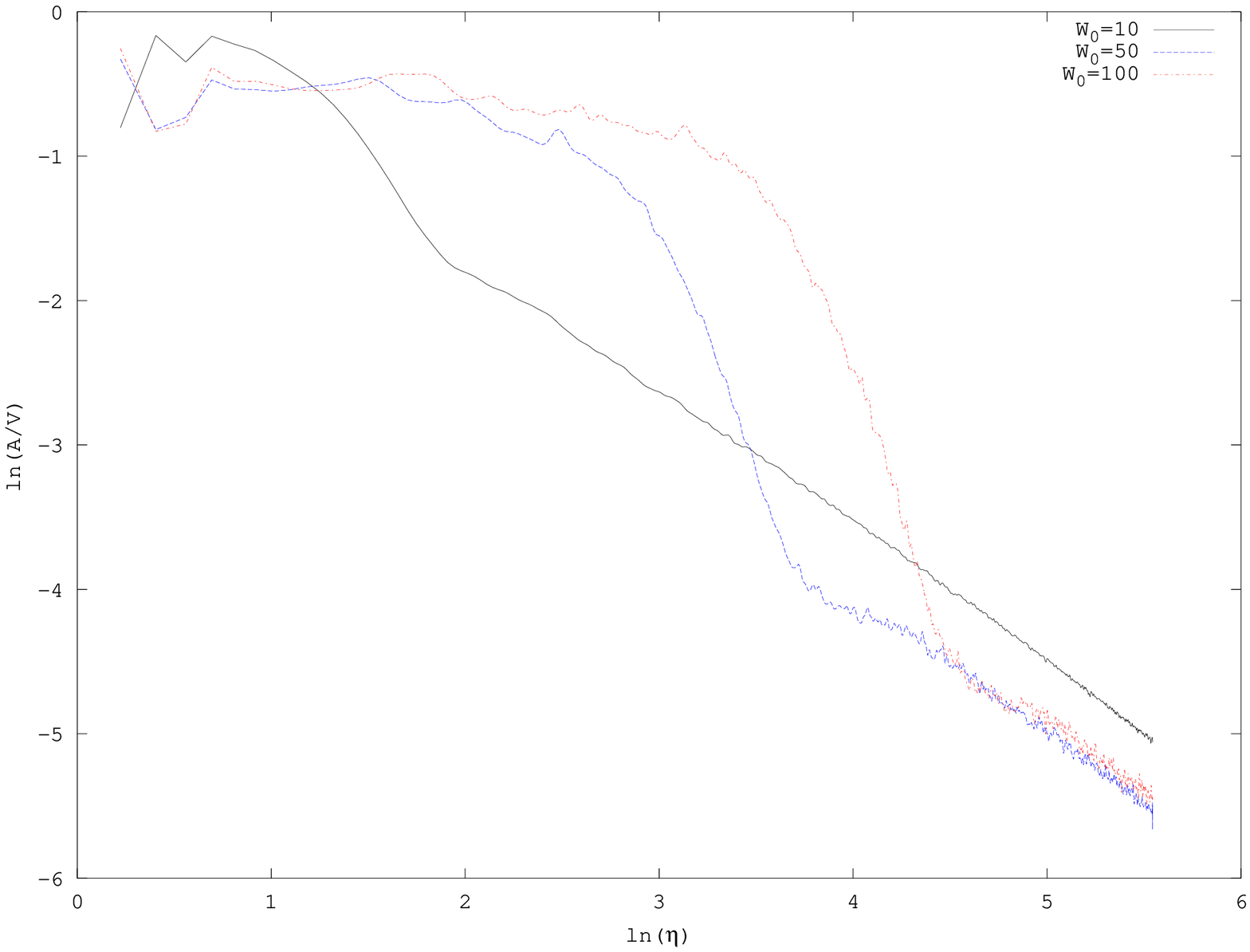}
\includegraphics[width=3.5in]{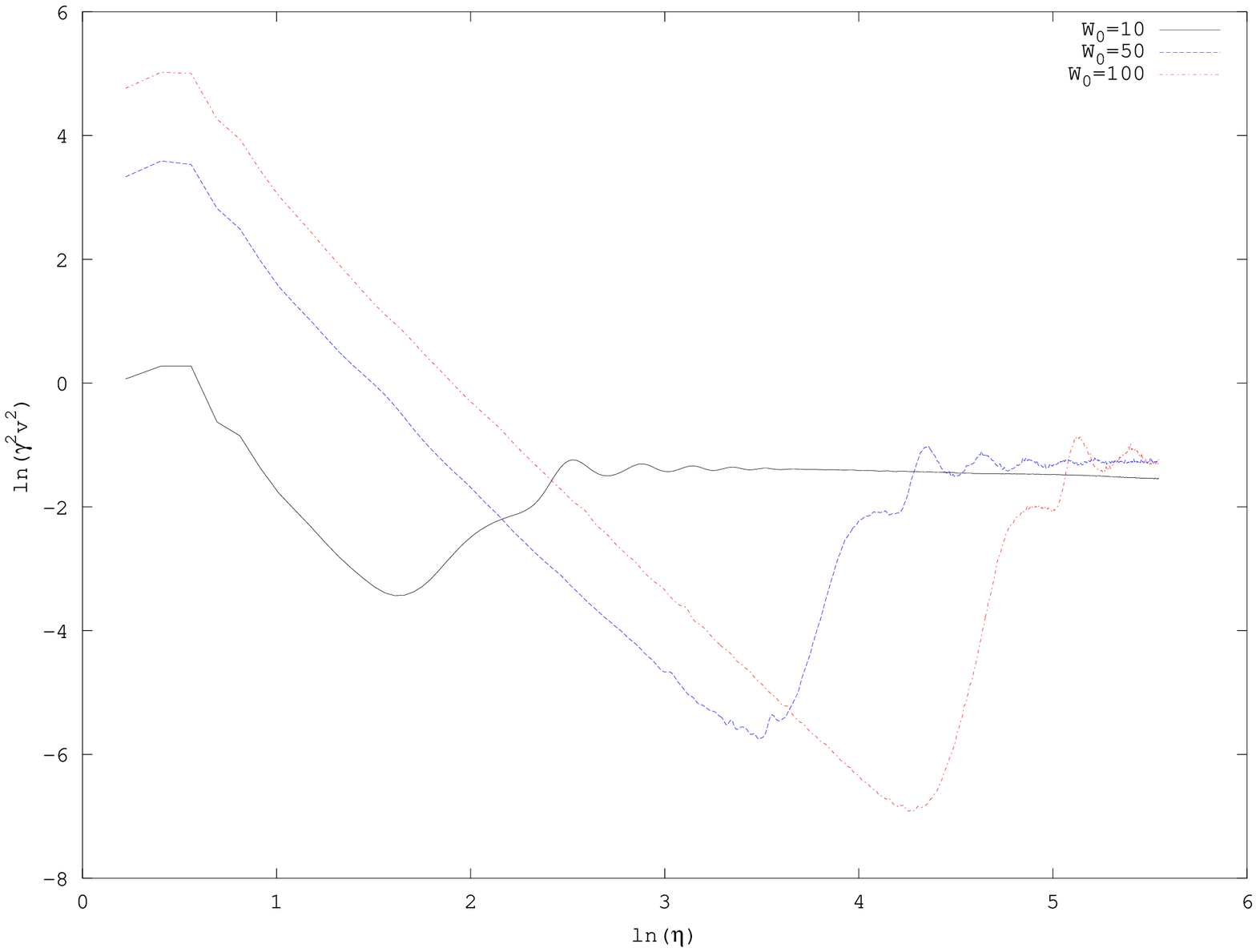}
\caption{\label{fig3}The evolution of the domain wall density ($\rho_w=A/V$) and velocity ($(\gamma v)^2$) as a function of conformal time, for the radiation era ($\lambda=1/2$) boxes of Table \protect\ref{tab2}, corresponding to different values of the (constant) wall thickness, $W_0=10,50,100$. The plotted curves are the average of 25 different simulations with different (random) initial conditions.}
\end{figure}

\begin{figure}
\includegraphics[width=3.5in]{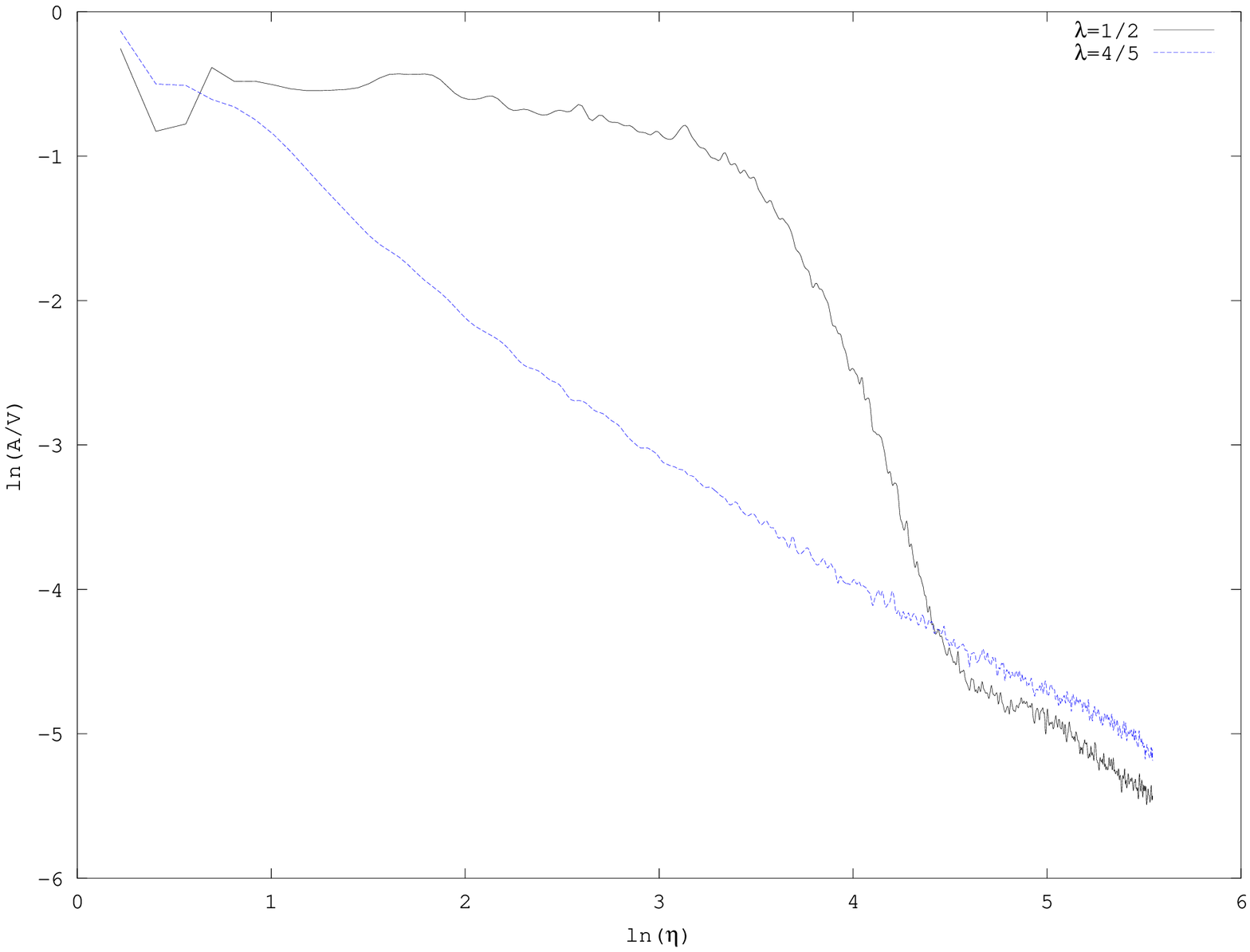}
\includegraphics[width=3.5in]{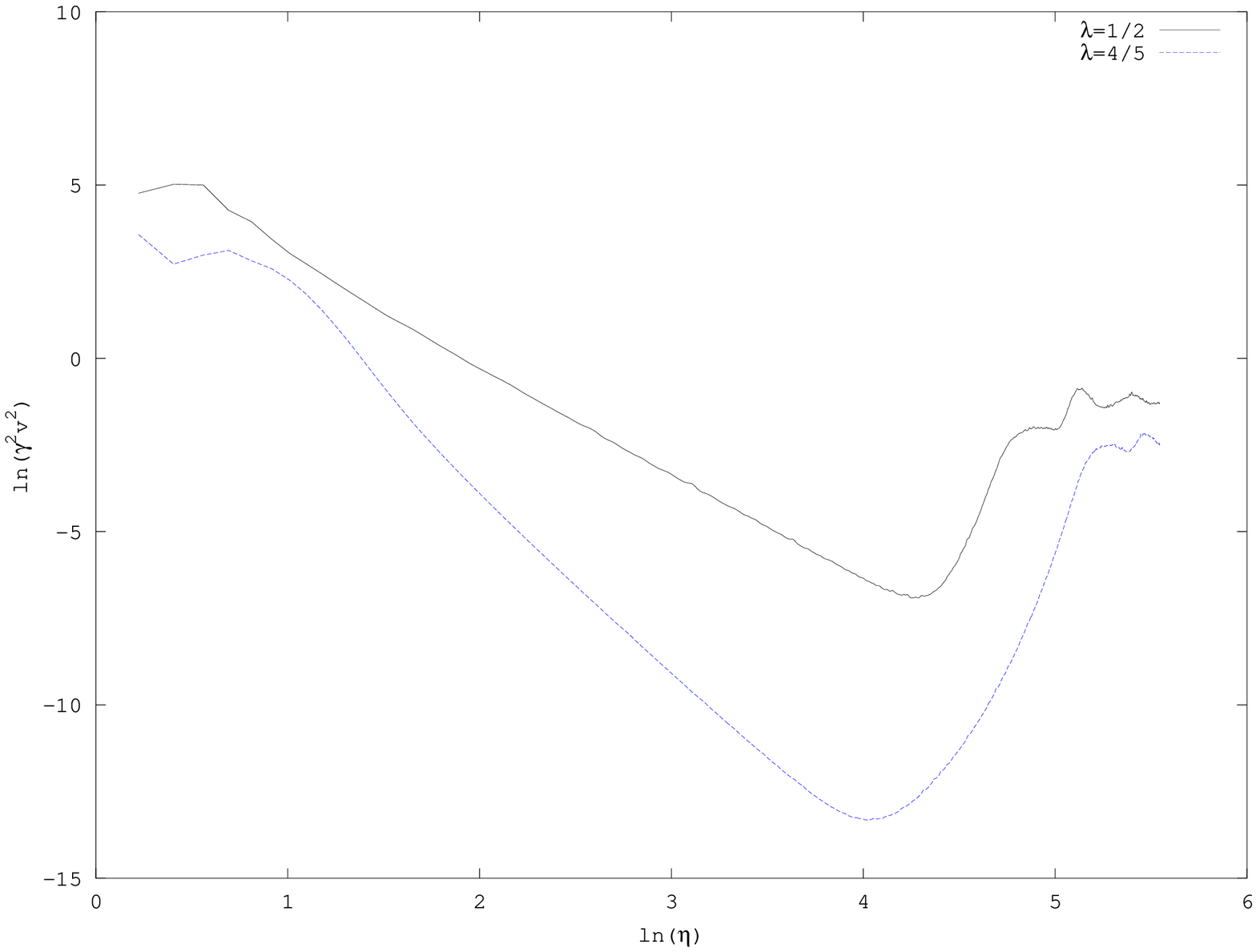}
\caption{\label{fig4}The evolution of the domain wall density ($\rho_w=A/V$) and velocity ($(\gamma v)^2$) as a function of conformal time, for the $W_0=100$ boxes of Table \protect\ref{tab2}, corresponding to different values of the expansion rate, $\lambda=1/2,4/5$. The plotted curves are the average of 25 different simulations with different (random) initial conditions.}
\end{figure}

Table \ref{tab2} and Figs. \ref{fig3} and \ref{fig4} show the compared results of radiation-era ($\lambda=1/2$) and fast expansion ($\lambda=4/5$) simulations, applying the PRS procedure to $512^3$ boxes and varying the thickness of the domain walls. As expected the relaxation time of the network is directly proportional to the wall thickness, being approximately given by the light crossing time of the walls.

Mindful of the fact that in the case of the thicker walls the network is just reaching the scaling solution at the end of the simulation, the late-time values of wall densities and velocities are consistent with each other within the numerical uncertainties. Note, in particular, that the velocities are remarkably similar; the differences in the wall densities may, at least in part, be due to the algorithm being used to identify the walls.

Bearing all this in mind, we believe that these results are consistent with the notion that the networks should eventually reach the same attractor solution regardless of the wall thickness. An example can be seen in Fig. \ref{fig3}, which shows the results of simulations with $W_0=10,50,100$ and all other parameters kept unchanged.

In practical terms one is therefore justified in using the smallest wall thickness compatible with an accurate identification of the walls. Previous results suggest \cite{Press} that this is around $W_0=10$, which is therefore the thickness used elsewhere in this paper. Nevertheless, similar results could have been obtained with a larger thickness, albeit with some additional computational cost (compare Fig. \ref{fig4} with Figs. \ref{fig1}, \ref{fig2} and \ref{fig5}).

\section{\label{num}Calibrating the VOS model}

\begin{figure}
\includegraphics[width=3.5in]{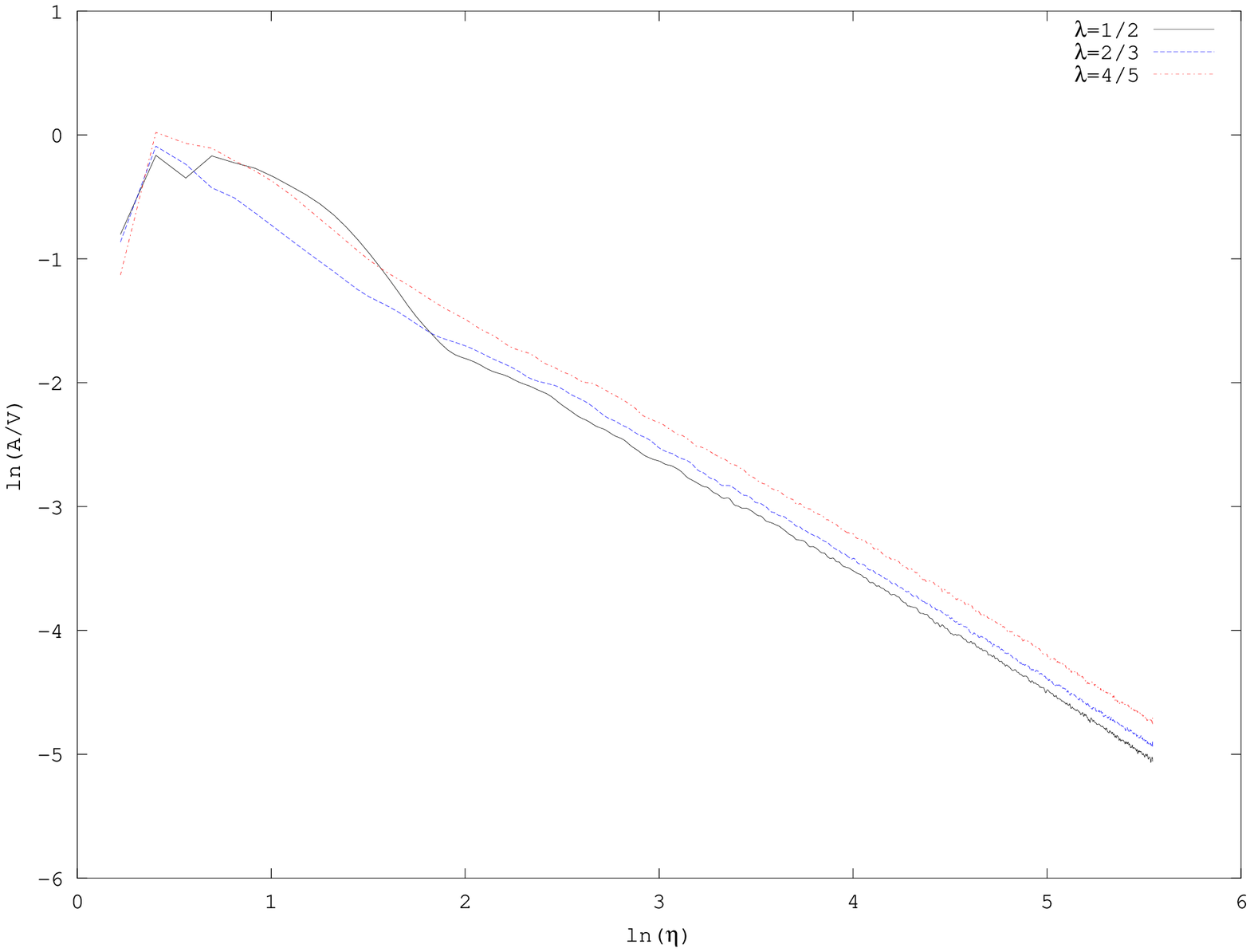}
\includegraphics[width=3.5in]{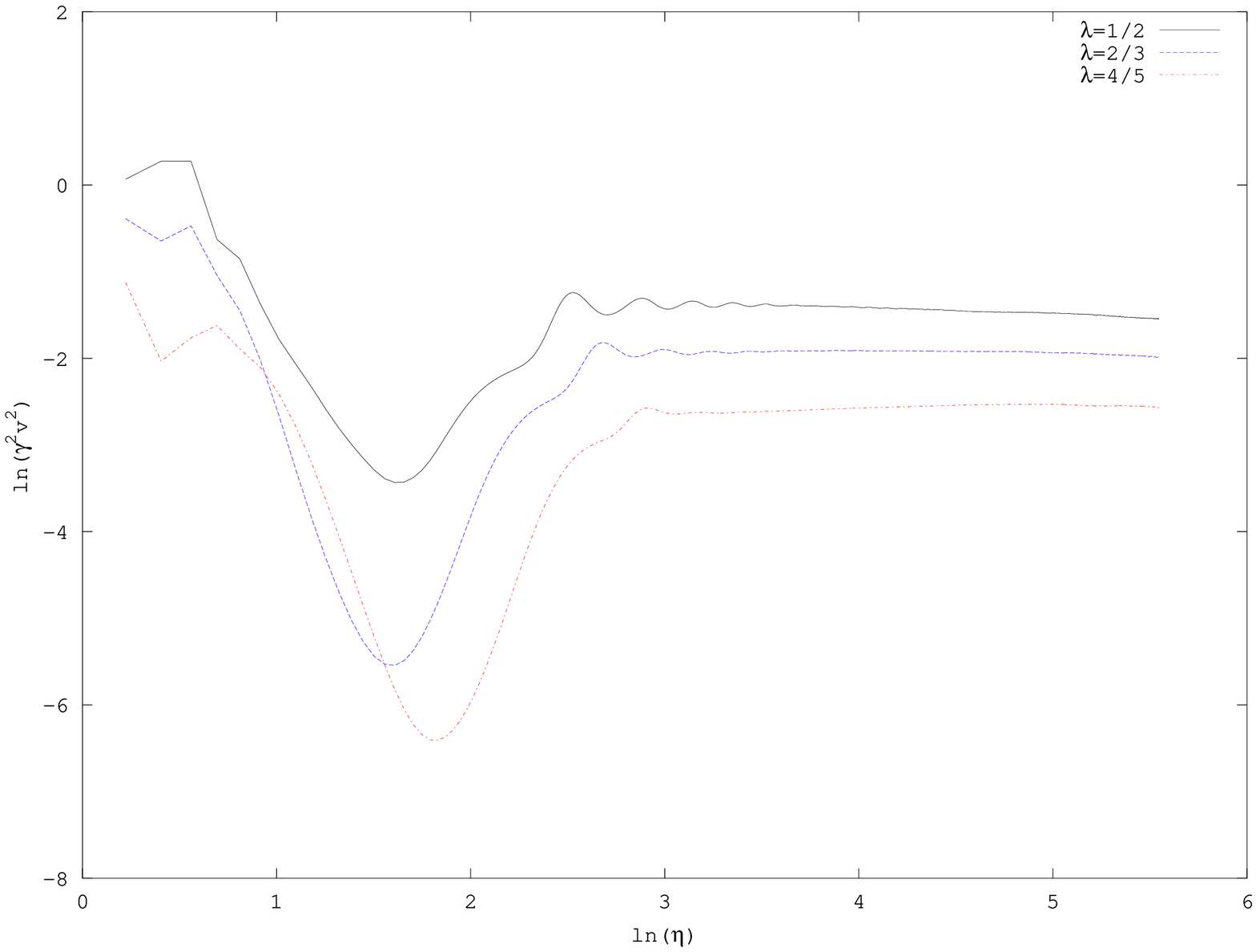}
\caption{\label{fig5}The evolution of the domain wall density ($\rho_w=A/V$) and velocity ($(\gamma v)^2$) as a function of conformal time, for the $512^3$ boxes of Table \protect\ref{tab2}, corresponding to different cosmological epochs: radiation era ($\lambda=1/2$), matter era ($\lambda=2/3$) and a faster expansion epoch with $\lambda=4/5$. The plotted curves are the average of 25 different simulations with different (random) initial conditions.}
\end{figure}

\begin{figure}
\includegraphics[width=3.5in]{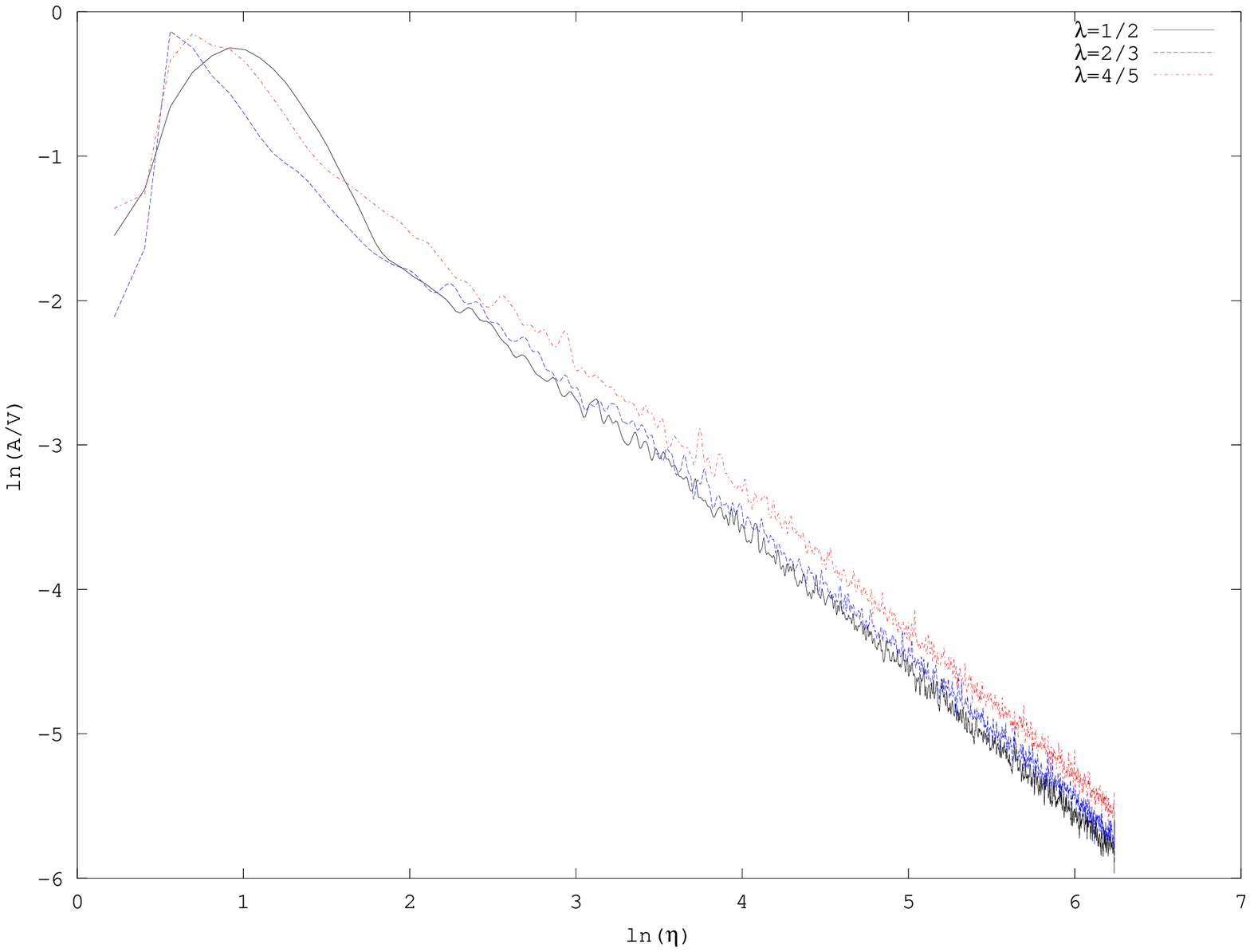}
\includegraphics[width=3.5in]{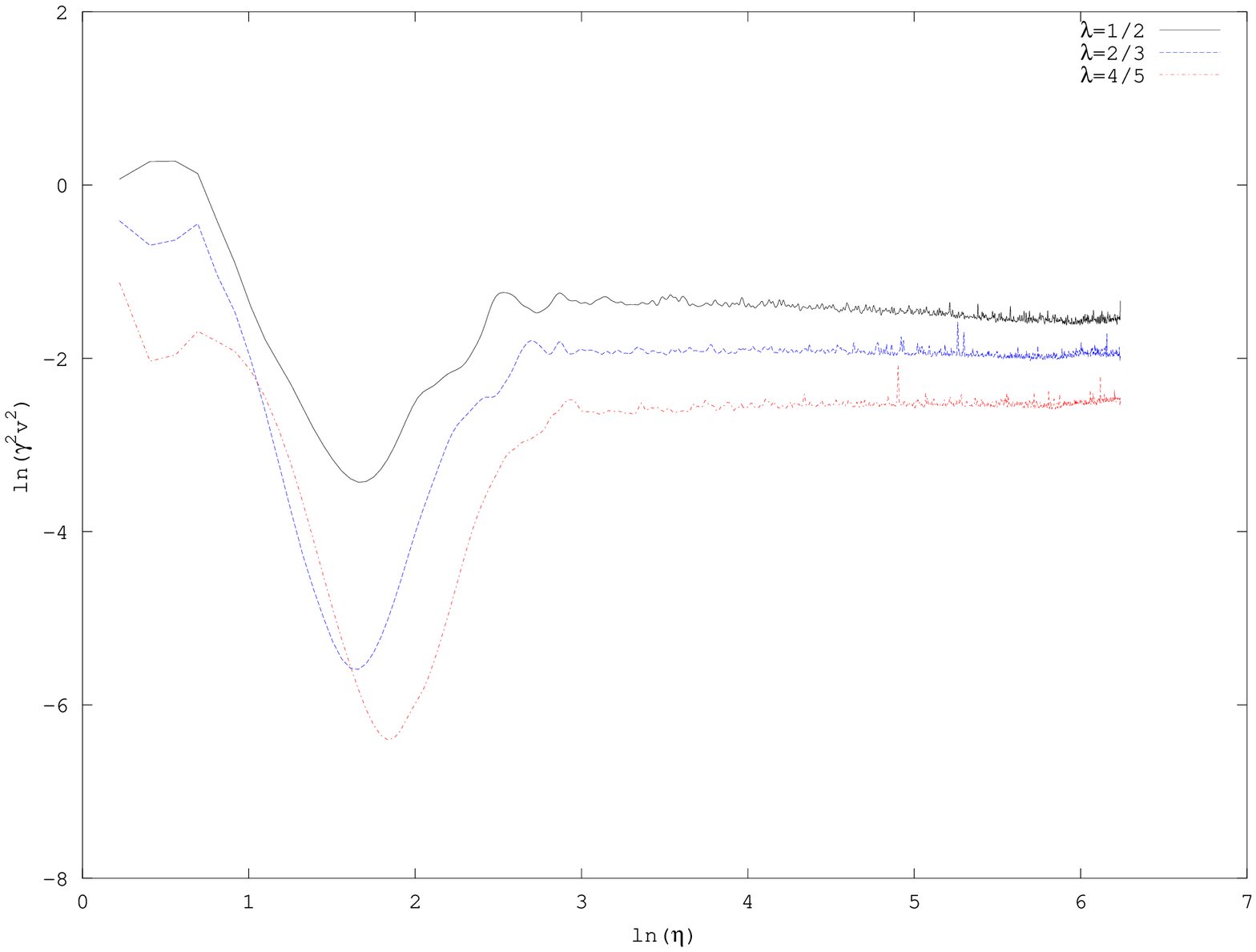}
\caption{\label{fig6}Same as Fig. \protect\ref{fig5}, for the $1024^3$ simulations.}
\end{figure}

Table \ref{tab3} and Figs. \ref{fig5} and \ref{fig6} show the compared results of very large 3D PRS simulations for three different cosmological expansion rates (parametrized by $\lambda$): the usual radiation and matter eras plus and additional fast expansion era (with $\lambda=4/5$). As was the case with the simulations described in the previous section, we find no deviation from the scaling behavior.

Moreover, comparing the scaling exponents for the correlation length in the $512^3$ and $1024^3$, and also for the $\alpha=3$, $256^3$ case of in Table \ref{tab1} one can observe that the exponents tend to become closer to $\mu=-1$ as we increase the box size. This further supports our suggestion, in the previous section, that deviations from scaling are likely to be due to the limited range of the simulations. A further contributing factor may the the fact that the initial conditions that are normally used in these simulations produce networks that are quite far from scaling, and therefore the numerical evolution of these networks will require a considerable dynamic range before scaling can be reached.

\begin{table*}
\begin{tabular}{|c|c|c|c|c|c|c|}
\hline 
Box Size&
$\lambda$&
Fit range ($\eta$)&
$\mu$&
$\nu$&
$\xi_c/\eta$&
$\gamma v$\tabularnewline
\hline 
\hline $512^3$ & 1/2 & 38.5-256 & $-0.99\pm0.05$ & $-0.0001\pm0.0002$ & $0.60\pm0.05 $ & $0.46\pm0.04 $ \tabularnewline
\hline $1024^3$ & 1/2 & 21-512 & $-0.99\pm0.05$ & $-0.0001\pm0.0001$ & $0.64\pm0.03 $ & $0.48\pm0.07 $ \tabularnewline
\hline
\hline $512^3$ & 2/3 & 26-256 & $-0.97\pm0.04$ & $-0.0000\pm0.0001$ & $0.54\pm0.01 $ & $0.37\pm0.02 $ 
\tabularnewline
\hline $1024^3$ & 2/3 & 21-512 & $-0.98\pm0.02$ & $-0.00001\pm0.00005$ & $0.62\pm0.03 $ & $0.37\pm0.03 $ \tabularnewline
\hline
\hline $512^3$ & 4/5 & 21-256 & $-0.96\pm0.03$ & $0.00001\pm0.00005$ & $0.44\pm0.01 $ & $0.28\pm0.02 $ \tabularnewline
\hline $1024^3$ & 4/5 & 26-512 & $-0.99\pm0.03$ & $0.00001\pm0.00003$ & $0.50\pm0.02 $ & $0.29\pm0.07 $ \tabularnewline
\hline
\end{tabular}
\caption{\label{tab3}The measured scaling properties of $3D$ PRS ($\alpha=3$, $\beta=0$) numerical simulations of domain wall networks with different box sizes and expansion rates ($\lambda$). The third column shows the range of the part of the simulation that was actually used in order to fit for the scaling exponents $\mu$ and $\nu$. The last two columns show the directly measured asymptotic values of $\xi_c/\eta$ and $\gamma v$, which can be related to the macroscopic parameters of the analytic model. All simulations have a constant wall thickness $W_0=10$ and all error bars are one-sigma statistical uncertainties obtained by averaging over 25 simulations with different random initial conditions.}
\end{table*}

\begin{table}
\begin{tabular}{|c|c|c|}
\hline 
$\lambda$&
$c_w$&
$k_w$\tabularnewline
\hline 
\hline 1/2 & $0.7\pm0.3 $ & $0.8\pm0.1 $ \tabularnewline
\hline 2/3 & $0.5\pm0.2 $ & $1.2\pm0.1 $ \tabularnewline
\hline 4/5 & $0.2\pm0.3 $ & $1.6\pm0.2 $ \tabularnewline
\hline
\hline Weighted mean & $0.5\pm0.2 $ & $1.1\pm0.3 $ \tabularnewline
\hline
\end{tabular}
\caption{\label{tab4}The calculated values of the VOS model parameters $c_w$ and $k_w$ for the $1024^3$ boxes of Table \protect\ref{tab3}. The final line show the best fit (weighted mean) result obtained by combining the values from the three cosmological epochs.}
\end{table}

Given the large size and dynamic range of these simulations, in this table we now show the values of $\xi_c/\eta$ and $\gamma v$ with error bars. As before we assume that the networks are scaling exactly, and the values are calculated from the last few timesteps in each simulation. We have experimented with several ways of calculating these values, for example first averaging the 25 simulations and then calculating the scaling values from the averaged run \textit{versus} calculating the scaling values from the individual simulations and then determining the average of these values, and found that provided one has a large enough sample (typically involving the last 100 timesteps of each simulation) the results are remarkably consistent. The uncertainties shown are purely the one-sigma statistical errors for each set of 25 simulations.

As expected, the results show that a faster expansion rate increases the damping on the walls, and therefore leads to smaller wall velocities and a larger wall density (which corresponds to a smaller correlation length). The agreement between the scaling values obtained from the $512^3$ and $1024^3$ is also quite encouraging, and suggests that we are indeed seeing the network's attractor scaling solution. Nevertheless, it should be pointed out that the measurements of the velocities are in closer agreement than those of the correlation length. This is likely to be due to the fact that our algorithm for identifying the walls and adding their areas is not completely accurate.

These results can now be used to calibrate the VOS walls model. Recall that in this model the scaling solution is parametrized by the expansion rate $\lambda$ (such that $a\propto t^\lambda$) and the phenomenological parameters $c_w$ and $k_w$. Given these parameters, the predicted values for the scaling arameters $\epsilon$ and $v$ are given by Eqs. (\ref{scaling1}--\ref{scaling2}). Using the results of Table \ref{tab3} we can
trivially obtain the value of $v$, and $\epsilon$ is also easily obtained from
\begin{equation}
\epsilon=\frac{L}{t}=\frac{\xi_c}{(1-\lambda)\eta}\,;
\label{findc}
\end{equation}
from these one finally obtains the numerically measured values of $c_w$ and $k_w$. The results of this analysis, using the data from the $1024^3$ runs, are shown in Table \ref{tab4}. We first calculated these values separately for each of the three cosmological epochs that we simulated, and we finally combined the three into a final, weighted mean set of results. Since these purely statistical errors may be underestimates, we rescaled the variance in the standard way, by multiplying it by the chi-squared per degree of freedom. Our final calibrated model parameters are
\begin{equation}
c_w=0.5\pm0.2\,
\label{newcw}
\end{equation}
and
\begin{equation}
k_w=1.1\pm0.3\,;
\label{newkw}
\end{equation}
these results are remarkably consistent with the previous, more qualitative analysis in \cite{MY2}, which had found $c_w\sim0.5$ and $k_m\sim0.9$.

\section{\label{conc}Conclusions}

We took advantage of recent improvements in computing power to numerically study the evolution of the simplest domain wall networks. Having carried out sets high-resolution, large dynamic range simulations with various amounts of damping, we find strong support for the suggestion that the attractor solution for the evolution of these networks is a linear scaling solution, with $\xi_{phys}\propto t$ (or equivalently $\xi_c\propto \eta$) and $v=const$. Our results suggest that previous hints of deviations from this behavior may have been due to the limited dynamical range of those simulations.

Moreover, we have used the results of the largest ($1024^3$) of our simulations to provide a calibration for the velocity-dependent one-scale model for domain walls. As a consistency check, we have used results from simulations in three different cosmological epochs, even though the model only has two free parameters, one of which quantifies the  network's energy loss rate while the other describes the (curvature-related) forces acting on the walls. Given the conceptual simplicity of the analytic model, we believe that the present numerical results support its validity, and suggest that it can be reliably used as a tool to study the cosmological consequences of these networks in quantitative detail.

This combination of analytical and numerical techniques, leading to a detailed calibration of a model (which so far had only been carried out for cosmic strings \cite{Moore,Unified}) can in principle be extended to networks of domain walls with junctions. The larger number of degrees of freedom (corresponding to additional scalar fields) makes the study of these models numerically trickier since in most cases the factor limiting the size of the boxes that can be simulated is memory rather than time, but otherwise our methods are directly appicable there. Another case of interest is that of semilocal string networks: here some steps towards an accurate model calibration have been taken recently \cite{Nunes}, and we shall return to it in a subsequent publication.

\section{acknowledgments}
The work of CM is funded by a Ci\^encia2007 Research Contract, funded by FCT/MCTES (Portugal) and POPH/FSE (EC), and we both acknowledge additional support from project PTDC/FIS/111725/2009 from FCT, Portugal.

Most of the numerical simulations in this paper were performed on the COSMOS Consortium supercomputer within the DiRAC Facility jointly funded by STFC and the Large Facilities Capital Fund of BIS (UK).
We are grateful to Andrey Kaliazin for his help with the code optimization.

%%%%%%%%%%%%%%%%%%%%%%%%%%%%%%%%%%%%%%%%%%%%%%%%%%%%%%%%%%%

\bibliography{walls}

\begin{thebibliography}{27}
\expandafter\ifx\csname natexlab\endcsname\relax\def\natexlab#1{#1}\fi
\expandafter\ifx\csname bibnamefont\endcsname\relax
  \def\bibnamefont#1{#1}\fi
\expandafter\ifx\csname bibfnamefont\endcsname\relax
  \def\bibfnamefont#1{#1}\fi
\expandafter\ifx\csname citenamefont\endcsname\relax
  \def\citenamefont#1{#1}\fi
\expandafter\ifx\csname url\endcsname\relax
  \def\url#1{\texttt{#1}}\fi
\expandafter\ifx\csname urlprefix\endcsname\relax\def\urlprefix{URL }\fi
\providecommand{\bibinfo}[2]{#2}
\providecommand{\eprint}[2][]{\url{#2}}

\bibitem[{\citenamefont{Kibble}(1976)}]{Kibble}
\bibinfo{author}{\bibfnamefont{T.~W.~B.} \bibnamefont{Kibble}},
  \bibinfo{journal}{J. Phys.} \textbf{\bibinfo{volume}{A9}},
  \bibinfo{pages}{1387} (\bibinfo{year}{1976}).

\bibitem[{\citenamefont{Vilenkin and Shellard}(1994)}]{Book}
\bibinfo{author}{\bibfnamefont{A.}~\bibnamefont{Vilenkin}} \bibnamefont{and}
  \bibinfo{author}{\bibfnamefont{E.~P.~S.} \bibnamefont{Shellard}}
  (\bibinfo{year}{1994}), \bibinfo{note}{{ }Cambridge, U.K.: Cambridge
  University Press}.

\bibitem[{\citenamefont{Zeldovich et~al.}(1974)\citenamefont{Zeldovich,
  Kobzarev, and Okun}}]{Zeldovich}
\bibinfo{author}{\bibfnamefont{Y.~B.} \bibnamefont{Zeldovich}},
  \bibinfo{author}{\bibfnamefont{I.~Y.} \bibnamefont{Kobzarev}},
  \bibnamefont{and} \bibinfo{author}{\bibfnamefont{L.~B.} \bibnamefont{Okun}},
  \bibinfo{journal}{Zh. Eksp. Teor. Fiz.} \textbf{\bibinfo{volume}{67}},
  \bibinfo{pages}{3} (\bibinfo{year}{1974}).

\bibitem[{\citenamefont{Avelino et~al.}(2008)\citenamefont{Avelino, Martins,
  Menezes, Menezes, and Oliveira}}]{Junctions}
\bibinfo{author}{\bibfnamefont{P.}~\bibnamefont{Avelino}},
  \bibinfo{author}{\bibfnamefont{C.}~\bibnamefont{Martins}},
  \bibinfo{author}{\bibfnamefont{J.}~\bibnamefont{Menezes}},
  \bibinfo{author}{\bibfnamefont{R.}~\bibnamefont{Menezes}}, \bibnamefont{and}
  \bibinfo{author}{\bibfnamefont{J.}~\bibnamefont{Oliveira}},
  \bibinfo{journal}{Phys.Rev.} \textbf{\bibinfo{volume}{D78}},
  \bibinfo{pages}{103508} (\bibinfo{year}{2008}), \eprint{0807.4442}.

\bibitem[{\citenamefont{Press et~al.}(1989)\citenamefont{Press, Ryden, and
  Spergel}}]{Press}
\bibinfo{author}{\bibfnamefont{W.~H.} \bibnamefont{Press}},
  \bibinfo{author}{\bibfnamefont{B.~S.} \bibnamefont{Ryden}}, \bibnamefont{and}
  \bibinfo{author}{\bibfnamefont{D.~N.} \bibnamefont{Spergel}},
  \bibinfo{journal}{Astrophys. J.} \textbf{\bibinfo{volume}{347}},
  \bibinfo{pages}{590} (\bibinfo{year}{1989}).

\bibitem[{\citenamefont{Dvali and Tye}(1999)}]{tye0}
\bibinfo{author}{\bibfnamefont{G.~R.} \bibnamefont{Dvali}} \bibnamefont{and}
  \bibinfo{author}{\bibfnamefont{S.~H.~H.} \bibnamefont{Tye}},
  \bibinfo{journal}{Phys. Lett.} \textbf{\bibinfo{volume}{B450}},
  \bibinfo{pages}{72} (\bibinfo{year}{1999}), \eprint{hep-ph/9812483}.

\bibitem[{\citenamefont{Brax and van~de Bruck}(2003)}]{Brax}
\bibinfo{author}{\bibfnamefont{P.}~\bibnamefont{Brax}} \bibnamefont{and}
  \bibinfo{author}{\bibfnamefont{C.}~\bibnamefont{van~de Bruck}},
  \bibinfo{journal}{Class. Quant. Grav.} \textbf{\bibinfo{volume}{20}},
  \bibinfo{pages}{R201} (\bibinfo{year}{2003}), \eprint{hep-th/0303095}.

\bibitem[{\citenamefont{Matsuda}(2004)}]{Matsuda}
\bibinfo{author}{\bibfnamefont{T.}~\bibnamefont{Matsuda}},
  \bibinfo{journal}{JHEP} \textbf{\bibinfo{volume}{10}}, \bibinfo{pages}{042}
  (\bibinfo{year}{2004}), \eprint{hep-ph/0406064}.

\bibitem[{\citenamefont{Barnaby et~al.}(2005)\citenamefont{Barnaby, Berndsen,
  Cline, and Stoica}}]{Barnaby}
\bibinfo{author}{\bibfnamefont{N.}~\bibnamefont{Barnaby}},
  \bibinfo{author}{\bibfnamefont{A.}~\bibnamefont{Berndsen}},
  \bibinfo{author}{\bibfnamefont{J.~M.} \bibnamefont{Cline}}, \bibnamefont{and}
  \bibinfo{author}{\bibfnamefont{H.}~\bibnamefont{Stoica}},
  \bibinfo{journal}{JHEP} \textbf{\bibinfo{volume}{06}}, \bibinfo{pages}{075}
  (\bibinfo{year}{2005}), \eprint{hep-th/0412095}.

\bibitem[{\citenamefont{Martins and Shellard}(1996{\natexlab{a}})}]{ms1a}
\bibinfo{author}{\bibfnamefont{C.~J. A.~P.} \bibnamefont{Martins}}
  \bibnamefont{and} \bibinfo{author}{\bibfnamefont{E.~P.~S.}
  \bibnamefont{Shellard}}, \bibinfo{journal}{Phys. Rev.}
  \textbf{\bibinfo{volume}{D53}}, \bibinfo{pages}{575}
  (\bibinfo{year}{1996}{\natexlab{a}}), \eprint{hep-ph/9507335}.

\bibitem[{\citenamefont{Martins and Shellard}(1996{\natexlab{b}})}]{ms1b}
\bibinfo{author}{\bibfnamefont{C.~J. A.~P.} \bibnamefont{Martins}}
  \bibnamefont{and} \bibinfo{author}{\bibfnamefont{E.~P.~S.}
  \bibnamefont{Shellard}}, \bibinfo{journal}{Phys. Rev.}
  \textbf{\bibinfo{volume}{D54}}, \bibinfo{pages}{2535}
  (\bibinfo{year}{1996}{\natexlab{b}}), \eprint{hep-ph/9602271}.

\bibitem[{\citenamefont{Martins and Shellard}(2002)}]{extend}
\bibinfo{author}{\bibfnamefont{C.~J. A.~P.} \bibnamefont{Martins}}
  \bibnamefont{and} \bibinfo{author}{\bibfnamefont{E.~P.~S.}
  \bibnamefont{Shellard}}, \bibinfo{journal}{Phys. Rev.}
  \textbf{\bibinfo{volume}{D65}}, \bibinfo{pages}{043514}
  (\bibinfo{year}{2002}), \eprint{hep-ph/0003298}.

\bibitem[{\citenamefont{Coulson et~al.}(1996)\citenamefont{Coulson, Lalak, and
  Ovrut}}]{Coulson}
\bibinfo{author}{\bibfnamefont{D.}~\bibnamefont{Coulson}},
  \bibinfo{author}{\bibfnamefont{Z.}~\bibnamefont{Lalak}}, \bibnamefont{and}
  \bibinfo{author}{\bibfnamefont{B.~A.} \bibnamefont{Ovrut}},
  \bibinfo{journal}{Phys. Rev.} \textbf{\bibinfo{volume}{D53}},
  \bibinfo{pages}{4237} (\bibinfo{year}{1996}).

\bibitem[{\citenamefont{Larsson et~al.}(1997)\citenamefont{Larsson, Sarkar, and
  White}}]{Larsson}
\bibinfo{author}{\bibfnamefont{S.~E.} \bibnamefont{Larsson}},
  \bibinfo{author}{\bibfnamefont{S.}~\bibnamefont{Sarkar}}, \bibnamefont{and}
  \bibinfo{author}{\bibfnamefont{P.~L.} \bibnamefont{White}},
  \bibinfo{journal}{Phys. Rev.} \textbf{\bibinfo{volume}{D55}},
  \bibinfo{pages}{5129} (\bibinfo{year}{1997}), \eprint{hep-ph/9608319}.

\bibitem[{\citenamefont{Avelino and Martins}(2000)}]{Fossils}
\bibinfo{author}{\bibfnamefont{P.~P.} \bibnamefont{Avelino}} \bibnamefont{and}
  \bibinfo{author}{\bibfnamefont{C.~J. A.~P.} \bibnamefont{Martins}},
  \bibinfo{journal}{Phys. Rev.} \textbf{\bibinfo{volume}{D62}},
  \bibinfo{pages}{103510} (\bibinfo{year}{2000}),
  \eprint[http://arXiv.org/abs]{astro-ph/0003231}.

\bibitem[{\citenamefont{Garagounis and Hindmarsh}(2003)}]{Garagounis}
\bibinfo{author}{\bibfnamefont{T.}~\bibnamefont{Garagounis}} \bibnamefont{and}
  \bibinfo{author}{\bibfnamefont{M.}~\bibnamefont{Hindmarsh}},
  \bibinfo{journal}{Phys. Rev.} \textbf{\bibinfo{volume}{D68}},
  \bibinfo{pages}{103506} (\bibinfo{year}{2003}), \eprint{hep-ph/0212359}.

\bibitem[{\citenamefont{Oliveira et~al.}(2005)\citenamefont{Oliveira, Martins,
  and Avelino}}]{MY1}
\bibinfo{author}{\bibfnamefont{J.}~\bibnamefont{Oliveira}},
  \bibinfo{author}{\bibfnamefont{C.}~\bibnamefont{Martins}}, \bibnamefont{and}
  \bibinfo{author}{\bibfnamefont{P.}~\bibnamefont{Avelino}},
  \bibinfo{journal}{Phys.Rev.} \textbf{\bibinfo{volume}{D71}},
  \bibinfo{pages}{083509} (\bibinfo{year}{2005}), \eprint{hep-ph/0410356}.

\bibitem[{\citenamefont{Avelino et~al.}(2005)\citenamefont{Avelino, Martins,
  and Oliveira}}]{MY2}
\bibinfo{author}{\bibfnamefont{P.}~\bibnamefont{Avelino}},
  \bibinfo{author}{\bibfnamefont{C.}~\bibnamefont{Martins}}, \bibnamefont{and}
  \bibinfo{author}{\bibfnamefont{J.}~\bibnamefont{Oliveira}},
  \bibinfo{journal}{Phys.Rev.} \textbf{\bibinfo{volume}{D72}},
  \bibinfo{pages}{083506} (\bibinfo{year}{2005}), \eprint{hep-ph/0507272}.

\bibitem[{\citenamefont{Lalak et~al.}(2008)\citenamefont{Lalak, Lola, and
  Magnowski}}]{Lalak}
\bibinfo{author}{\bibfnamefont{Z.}~\bibnamefont{Lalak}},
  \bibinfo{author}{\bibfnamefont{S.}~\bibnamefont{Lola}}, \bibnamefont{and}
  \bibinfo{author}{\bibfnamefont{P.}~\bibnamefont{Magnowski}},
  \bibinfo{journal}{Phys.Rev.} \textbf{\bibinfo{volume}{D78}},
  \bibinfo{pages}{085020} (\bibinfo{year}{2008}), \eprint{0710.1233}.

\bibitem[{\citenamefont{Hindmarsh}(2003)}]{Hindmarsh}
\bibinfo{author}{\bibfnamefont{M.}~\bibnamefont{Hindmarsh}},
  \bibinfo{journal}{Phys. Rev.} \textbf{\bibinfo{volume}{D68}},
  \bibinfo{pages}{043510} (\bibinfo{year}{2003}), \eprint{hep-ph/0207267}.

\bibitem[{\citenamefont{Moore et~al.}(2002)\citenamefont{Moore, Shellard, and
  Martins}}]{Moore}
\bibinfo{author}{\bibfnamefont{J.~N.} \bibnamefont{Moore}},
  \bibinfo{author}{\bibfnamefont{E.~P.~S.} \bibnamefont{Shellard}},
  \bibnamefont{and} \bibinfo{author}{\bibfnamefont{C.~J. A.~P.}
  \bibnamefont{Martins}}, \bibinfo{journal}{Phys. Rev.}
  \textbf{\bibinfo{volume}{D65}}, \bibinfo{pages}{023503}
  (\bibinfo{year}{2002}), \eprint[http://arXiv.org/abs]{hep-ph/0107171}.

\bibitem[{\citenamefont{Sousa and Avelino}(2010)}]{Sousa}
\bibinfo{author}{\bibfnamefont{L.}~\bibnamefont{Sousa}} \bibnamefont{and}
  \bibinfo{author}{\bibfnamefont{P.}~\bibnamefont{Avelino}},
  \bibinfo{journal}{Phys.Rev.} \textbf{\bibinfo{volume}{D81}},
  \bibinfo{pages}{087305} (\bibinfo{year}{2010}), \eprint{1101.3350}.

\bibitem[{\citenamefont{Martins and Achucarro}(2008)}]{Monopoles}
\bibinfo{author}{\bibfnamefont{C.}~\bibnamefont{Martins}} \bibnamefont{and}
  \bibinfo{author}{\bibfnamefont{A.}~\bibnamefont{Achucarro}},
  \bibinfo{journal}{Phys.Rev.} \textbf{\bibinfo{volume}{D78}},
  \bibinfo{pages}{083541} (\bibinfo{year}{2008}).

\bibitem[{\citenamefont{Kawano}(1990)}]{kawano}
\bibinfo{author}{\bibfnamefont{L.}~\bibnamefont{Kawano}},
  \bibinfo{journal}{Phys. Rev.} \textbf{\bibinfo{volume}{D41}},
  \bibinfo{pages}{1013} (\bibinfo{year}{1990}).

\bibitem[{\citenamefont{Martins}(2004)}]{nonint}
\bibinfo{author}{\bibfnamefont{C.~J. A.~P.} \bibnamefont{Martins}},
  \bibinfo{journal}{Phys. Rev.} \textbf{\bibinfo{volume}{D70}},
  \bibinfo{pages}{107302} (\bibinfo{year}{2004}), \eprint{hep-ph/0410326}.

\bibitem[{\citenamefont{Martins et~al.}(2004)\citenamefont{Martins, Moore, and
  Shellard}}]{Unified}
\bibinfo{author}{\bibfnamefont{C.}~\bibnamefont{Martins}},
  \bibinfo{author}{\bibfnamefont{J.}~\bibnamefont{Moore}}, \bibnamefont{and}
  \bibinfo{author}{\bibfnamefont{E.}~\bibnamefont{Shellard}},
  \bibinfo{journal}{Phys.Rev.Lett.} \textbf{\bibinfo{volume}{92}},
  \bibinfo{pages}{251601} (\bibinfo{year}{2004}), \eprint{hep-ph/0310255}.

\bibitem[{\citenamefont{Nunes et~al.}(2011)\citenamefont{Nunes, Avgoustidis,
  Martins, and Urrestilla}}]{Nunes}
\bibinfo{author}{\bibfnamefont{A.}~\bibnamefont{Nunes}},
  \bibinfo{author}{\bibfnamefont{A.}~\bibnamefont{Avgoustidis}},
  \bibinfo{author}{\bibfnamefont{C.}~\bibnamefont{Martins}}, \bibnamefont{and}
  \bibinfo{author}{\bibfnamefont{J.}~\bibnamefont{Urrestilla}},
  \bibinfo{journal}{Phys.Rev.} \textbf{\bibinfo{volume}{D84}},
  \bibinfo{pages}{063504} (\bibinfo{year}{2011}), \eprint{1107.2008}.

\end{thebibliography}

\end{document}